\documentclass[aps,showpacs,amsmath,amssymb,superscriptaddress,prl,preprint]{revtex4-1}

\usepackage{graphicx}
\usepackage{bm}
\usepackage[dvips]{color} 

\begin{document}


\title{Rashba-Edelstein Magnetoresistance in Metallic Heterostructure}

\author{Hiroyasu Nakayama}
\affiliation{Department of Applied Physics and Physico-Informatics, Keio University, Yokohama 223-8522, Japan}

\author{Yusuke Kanno}
\affiliation{Department of Applied Physics and Physico-Informatics, Keio University, Yokohama 223-8522, Japan}

\author{Hongyu An}
\affiliation{Department of Applied Physics and Physico-Informatics, Keio University, Yokohama 223-8522, Japan}

\author{Takaharu Tashiro}
\affiliation{Department of Applied Physics and Physico-Informatics, Keio University, Yokohama 223-8522, Japan}

\author{Satoshi Haku}
\affiliation{Department of Applied Physics and Physico-Informatics, Keio University, Yokohama 223-8522, Japan}

\author{Akiyo Nomura}
\affiliation{Department of Applied Physics and Physico-Informatics, Keio University, Yokohama 223-8522, Japan}

\author{Kazuya Ando}
\email{ando@appi.keio.ac.jp}
\affiliation{Department of Applied Physics and Physico-Informatics, Keio University, Yokohama 223-8522, Japan}
\affiliation{PRESTO, Japan Science and Technology Agency, Kawaguchi, Saitama 332-0012, Japan}

\date{\today}


\begin{abstract}
We report the observation of magnetoresistance originating from Rashba spin-orbit coupling (SOC) in a metallic heterostructure: the Rashba-Edelstein (RE) magnetoresistance. We show that the simultaneous action of the direct and inverse RE effects in a Bi/Ag/CoFeB trilayer couples current-induced spin accumulation to the electric resistance. The electric resistance changes with the magnetic-field angle, reminiscent of the spin Hall magnetoresistance, despite the fact that bulk SOC is not responsible for the magnetoresistance. We further found that, even when the magnetization is saturated, the resistance increases with increasing the magnetic-field strength, which is attributed to the Hanle magnetoresistance in this system. 
\end{abstract}


\maketitle

The change of electrical resistance of ferromagnetic films and multilayers in a magnetic field, magnetoresistance, has been studied for a long time, providing fundamental understanding of spin-dependent transport in solids~\cite{Maekawa2}. Recently, magnetoresistance due to a nonequilibrium proximity effect has been observed in a heavy metal/magnetic insulator bilayer, where no charge current flows in the magnetic layer~\cite{Nakayama:206601}. This magnetoresistance is commonly referred to as the spin Hall magnetoresistance (SMR)~\cite{Nakayama:206601,PhysRevLett.116.097201,chen2013theory,althammer2013quantitative,vlietstra2013spin,hahn2013comparative}.

The physics behind the SMR is the spin-current reflection and the reciprocal spin-charge conversion caused by the simultaneous action of the spin Hall effect (SHE)~\cite{Dyakonov,Hirsch,Murakami,Sinova,Kato,Wunderlich,RevModPhys.87.1213,6516040} and inverse spin Hall effect (ISHE)~\cite{Valenzuela,KimuraPRL,Saitoh}. Such a situation can be realized even in the absence of the bulk spin-orbit coupling. In this Letter, we demonstrate magnetoresistance induced by interfacial spin-orbit coupling and spin-current reflection in a metallic heterostructure: the Rashba-Edelstein magnetoresistance (REMR). As shown in Figs.~1(a) and 1(b), a charge current carried by a two-dimensional electron gas with helical spin polarization is accompanied by a nonzero spin accumulation with the spins along the in-plane direction transverse to the applied charge current, which is known as the Rashba-Edelstein effect (REE)~\cite{edelstein1990spin,Sanchez,Sangiao,Nomura:212403,PhysRevB.93.2244192,Isasa:014420,Zhang:014420}. We show that the nonequilibrium spin accumulation created by the REE is coupled to the electric resistance in a Bi/Ag/CoFeB trilayer. The spin accumulation generated from a 2D charge current through the REE at the Bi/Ag interface diffuses as a 3D spin current ${\bf j}_\text{s}^\text{REE}$ in the Ag layer as shown in Fig.~1(c). This spin current is reflected at the Ag/CoFeB interface and the reflected 3D spin current ${\bf j}_\text{s}^\text{back}$ is then converted into a 2D charge current ${\bf j}_\text{c}^\text{IREE}$ through the inverse Rashba-Edelstein effect (IREE) [see Fig.~1(d)]. Thus, the simultaneous action of the REE and IREE gives rise to an additional charge current, or changes the electric resistance of the trilayer, which is the REMR.

The magnetoresistance due to the spin-current reflection is associated with the spin-current absorption into the ferromagnetic layer~\cite{Nakayama:206601}. Thus, to test the possibility of the existence of the REMR, we first quantify an anti-damping spin-orbit torque~\cite{Kim:240,Hayashi:144425,Yang:04EM06,Emori:611,Garello:587}, which is accompanied by the spin-current absorption, arising from the REE at the Bi/Ag interface. To measure the anti-damping torque, we prepared a Bi/Ag/Pt/Co/Pt multilayer with perpendicular magnetic anisotropy shown in Fig.~2(a). The Bi/Ag/Pt/Co/Pt multilayer was deposited on a Gd$_{3}$Ga$_{5}$O$_{12}$ (GGG) (111) single crystalline substrate at room temperature by RF magnetron sputtering with a base pressure of around $3 \times 10^{-6}$ Pa. The stacking order of the multilayer is Bi/Ag/Pt/Co/Pt/GGG, and thus the Co layer is protected from oxidization. The multilayer was patterned into a Hall bar structure with the width of 500 $\mu$m. For the Bi/Ag/Pt/Co/Pt multilayer, we measured harmonic Hall voltages using two lock-in amplifiers with an applied charge current frequency of 507.32 Hz~\cite{Hayashi:144425}. The charge current was applied along the $x$-axis [see Fig.~2(a)]. All the measurements were performed at room temperature.

Figures~2(b) and 2(c) show the first $V_\omega$ and second $V_{2\omega}$ harmonic voltages for the Bi(5.0 nm)/Ag(3.0 nm)/Pt(3.0 nm)/Co(1.1 nm)/Pt(3.0 nm) multilayer measured when the external magnetic field $H_x$ was applied along the $x$-axis, parallel to the charge current. The applied charge current density for the multilayer was $j _\text{c}= 5.0 \times 10^{5}$ A/cm$^{2}$. The variation of $V_\omega$ shown in Fig.~2(b) is due to the rotation of the magnetization with $H_x$, which can be approximated by a quadratic function around $H_x=0$. Importantly, we observed nonzero second harmonic Hall voltage $V_{2\omega}$ for the Bi/Ag/Pt/Co/Pt multilayer as shown in Fig.~2(b), which demonstrates that a nonzero damping-like torque is generated in the multilayer. Here, the damping-like $H_\text{D}$ and field-like $H_\text{F}$ effective magnetic fields can be calculated using the following equation in which the contribution of the planar Hall effect (PHE) is included~\cite{Kim:240,Hayashi:144425,Yang:04EM06}: 
\begin{equation}
H_{\mathrm{D}\left(\mathrm{F}\right)}=\left(H'_{\mathrm{D}\left(\mathrm{F}\right)}\pm2\xi H'_{\mathrm{\mathrm{F}}\left(\mathrm{D}\right)}\right)  /\left(1-4\xi^{2}\right), \label{eq1}
\end{equation}
where $H'_{\mathrm{D}\left(\mathrm{F}\right)}=-2\left( {\partial V_{2\omega}}/{\partial H_{x\left(y\right)}}\right)/\left( {\partial^{2}V_{\omega}}/{\partial H_{x\left(y\right)}^{2}}\right) $ and $\xi=\Delta R_{\mathrm{PH}}/\Delta R_{\mathrm{AH}}$. Here, $H_{x(y)}$ is the magnetic field applied along the $x(y)$ direction. $\Delta R_{\mathrm{AH}}$ and $\Delta R_{\mathrm{PH}}$ are the Hall resistances due to the AHE and PHE, respectively. By fitting $V_\omega$ with a quadratic function and $V_{2\omega}$ with a linear function and using Eq. (\ref{eq1}) with the measured value of $\xi= 0.14$ for the multilayer, we obtain the current-induced damping-like effective field $\mu_0 H_\text{D}=0.06$ mT and field-like effective field $\mu_0 H_\text{F}=0.02$ mT. We have confirmed that the effective field is proportional to the applied current density [see Fig.~S1(h) in Supplementary Materials], consistent with previous reports~\cite{Emori:611,Kim:240}.

The observed damping-like torque generated in the Bi/Ag/Pt/Co/Pt multilayer originates from the REE at the Bi/Ag interface. We have confirmed that both the damping-like and field-like spin-orbit torques in a symmetric Pt(3.5 nm)/Co(1.1 nm)/Pt(3.5 nm) film are negligible due to the cancelation of the Rashba and spin Hall torques generated in the top Pt/Co and bottom Co/Pt junctions~\cite{Haney:174411,An:16XX,supplementary}. Furthermore, as shown in Figs.~2(c) and 2(d), we found that $V_{2\omega}$ is vanishingly small for a Ag(3.0 nm)/Pt(3.0 nm)/Co(1.1 nm)/Pt(3.0 nm) multilayer, in which the Bi layer is absent. This result shows that the damping-like torque due to the SHE in the Ag layer is negligible in the Bi/Ag/Pt/Co/Pt multilayer because of the small spin Hall angle $\theta_\text{SH}^\text{Ag} = 0.0068$~\cite{Wang:197201}. The SHE in the Bi layer is also unlikely to be the origin of the damping-like torque, since only less than 1\% of the applied current flows in the Bi layer because of its high resistivity. We can also rule out the damping-like torque that could be generated by the SHE in a thin layer of AgBi alloy because the sign of $V_{2\omega}$ observed for the Bi/Ag/Pt/Co/Pt multilayer is opposite to the prediction of the SHE of AgBi alloy. Therefore, our experimental observation, the nonzero damping-like torque in the Bi/Ag/Pt/Co/Pt multilayer and negligible damping-like torque in the Ag/Pt/Co/Pt multilayer, shows that the origin of the damping-like torque can be attributed to the diffusive spin current generated by the REE at the Bi/Ag interface. 

The damping-like spin-torque generation efficiency in the Bi/Ag/Pt/Co/Pt multilayer obtained from the above measurement, $\mu_0 H_\text{D}/j_\text{c}=1.2\times 10^{-7}$ mT/Acm$^{-2}$, is an order of magnitude smaller than that in heavy metal/ferromagnetic metal heterostructures~\cite{Emori:611}, for which the SMR has been observed~\cite{PhysRevLett.116.097201}. This low efficiency is due to the spin-current decay in the Pt layer; the 3D spin current created by the diffusion of the nonequilibrium spin accumulation generated at the Bi/Ag interface decays quickly in the Pt layer, which is roughly estimated as $j_\text{s}^\text{Pt/Co}/j_\text{s}^\text{Ag/Pt}\sim\exp(-d_\text{Pt}/\lambda_\text{Pt})\sim 0.14$, where $j_\text{s}^\text{Pt/Co(Ag/Pt)}$ is the spin current density at the Pt/Co(Ag/Pt) interface, $d_\text{Pt}=3$ nm and $\lambda_\text{Pt}=1.5 $ nm are the thickness and spin diffusion length of the Pt layer. The spin memory loss at the interfaces further suppresses the spin current density injected into the Co layer~\cite{supplementary,Sanchez:106602}. For the Hall-voltage measurements discussed above, the Co layer was sandwiched by the Pt layers to induce perpendicular magnetic anisotropy (PMA). However, the PMA is not necessary for the REMR. Thus, next we measure the magnetoresistance by fabricating a direct contact between the Bi/Ag junction and ferromagnetic layer not to reduce the spin current density arriving at the ferromagnetic layer.

We measured the longitudinal electric resistance $R$ of Bi(5 nm)/Ag($t_\text{Ag}$)/CoFeB(2.5 nm) trilayers with different Ag-layer thickness $t_\text{Ag}$. The stacking order of the device is Bi/Ag/CoFeB/GGG, and thus the CoFeB layer is protected from oxidization. Figure~3(a) shows the resistance of a Bi(5 nm)/CoFeB(2.5 nm) bilayer, i.e. $t_\text{Ag}=0$, during rotation of the applied magnetic field $\mu_0 H=10$ T in the $xy$, $zy$ and $zx$ planes [see Fig.~3(b)]. At $\mu_0 H=10$ T, the magnetization ${\bf M}$ of the CoFeB layer is saturated and follows the direction of ${\bf H}$. The rotation angles ($\alpha$, $\beta$, and $\gamma$) are defined in Fig.~3(b). As shown in Fig.~3(a), we observed a sizable magnetoresistance in all three orthogonal planes. The observed field-angle-dependent magnetoresistance (ADMR) is consistent with the prediction of the anisotropic magnetoresistance (AMR) of the CoFeB layer. Because the electrical resistance of the Bi layer is much larger than that of the CoFeB layer, the applied charge current flows mostly in the CoFeB layer. The AMR phenomenology of polycrystalline ferromagnetic metals predicts $\rho=\rho_\perp+\Delta\rho_\text{A} m_x^2$, where $\rho$ is the electric resistivity along the $x$ direction. $m_x$ is the $x$ component of ${\bf m}={\bf M}/M_\text{s}$, where $\mu_0 M_\text{s}$ is the saturation magnetization. $\Delta \rho_\text{A}$ ($=\rho_\parallel -\rho_\perp$) is the magnitude of the resistivity change as a function of the magnetization orientation, where $\rho_\parallel$ and $\rho_\perp$ are the resistivities for magnetizations aligned along and perpendicular to the applied charge current, respectively. This predicts the resistance change as $\Delta R(\alpha)\sim \cos^2\alpha$ and $\Delta R(\gamma)\sim \sin^2\gamma$ [see Fig.~3(a)]. In contrast, although the AMR predicts $\Delta R(\beta)=0$, we found $\Delta R(\beta)\sim \sin^2\beta$, which can be attributed to the geometrical size effect of the AMR~\cite{gil2005magnetoresistance}.

The symmetry of the ADMR changes clearly by inserting a Ag layer between the Bi and CoFeB layers. For a Bi(5 nm)/Ag(2 nm)/CoFeB(2.5 nm) trilayer, we observed $\Delta R(\beta)\sim -\sin^2\beta$ as shown in Fig.~3(c); $\Delta R(\beta)$ changes the sign by inserting the Ag layer [see Figs.~3(a) and 3(c)]. This is evidence of the REMR in the Bi/Ag/CoFeB trilayer. Since the symmetry of the REE, i.e. the spin-polarization of the 3D diffusive spin current is perpendicular to both the applied charge current and the flow direction of the spin current, is the same as that of the SHE, the field-angle dependence of the REMR is the same as that of SMR. Since the SMR resistivity can be formulated as $\rho=\rho_0-\Delta\rho _\text{S} m_y^2$~\cite{Nakayama:206601}, the SMR and REMR predict $\Delta R(\beta)\sim -\sin^2\beta$; the sign of $\Delta R(\beta)$ due to the REMR is opposite to that due to the AMR. Here $\rho_{0}$ is a constant resistivity offset, $m_y$ is the $y$ component of ${\bf m}$, and $\Delta \rho_\text{S}$ is the magnitude of the resistivity change due to the SMR or REMR. Therefore, the sign reversal of the ADMR due to the Ag-layer insertion is consistent with the prediction of the REMR in the Bi/Ag/CoFeB trilayer. The observed magnetoresistance cannot be attributed to the SMR due to the SHE in the Ag layer or the magnetoresistance due to the magnetic proximity effect at the Ag/CoFeB interface, since the sign reversal of $\Delta R(\beta)$ is absent in a Ag(2 nm)/CoFeB(2.5 nm) bilayer as shown in Fig.~3(d). In the Ag/CoFeB bilayer, the applied charge current flows mostly in the Ag layer due to the high electrical conductivity of the Ag layer. However, the SMR due to the SHE in the Ag layer is negligible because of the small spin Hall angle of Ag; the magnitude of the SMR is proportional to the square of the spin Hall angle~\cite{Nakayama:206601}. Thus, the magnetoresistance in the Ag/CoFeB bilayer is attributed to the geometrical size effect of the AMR in the CoFeB layer, as with the Bi/CoFeB bilayer. In contrast, in the Bi/Ag/CoFeB trilayer, the ADMR is dominated by the REMR, whose sign of $\Delta R(\beta)$ is opposite to that of the magnetoresistance in the CoFeB layer. This is the origin of the reversed $\Delta R(\beta)$ induced by inserting the Ag layer between the Bi and CoFeB layers. The $\alpha$-scan data shown in Fig.~3(c) is consistent with a previous result for a Ni$_{80}$Fe$_{10}$/Ag/Bi film~\cite{PhysRevB.93.2244192}.

To further verify the role of the Rashba effect at the Bi/Ag interface in the magnetoresistance, we examined Ag-layer thickness $t_\text{Ag}$ dependence of the magnetoresistance. Figure~3(e) shows $\Delta R/R$ as a function of $t_\text{Ag}$, where $\Delta R\equiv R(\beta=0)-R(\beta=90^\circ)$. This result is consistent with the prediction of the REMR in presence of the shunting current. In the trilayers with $t_\text{Ag}\sim0$, most of the applied charge current flows in the CoFeB layer and $\Delta R$ is dominated by the  change of the resistance $R_\text{CFB}$ of the CoFeB layer $\Delta R_\text{CFB}=R_\text{CFB}(\beta=0)-R_\text{CFB}(\beta=90^\circ)$. However, by increasing $t_\text{Ag}$, the Bi/Ag interface layer is formed, and therefore the REMR arising from the 2D charge current at the Bi/Ag interface emerges, changing the sign of the magnetoresistance of the trilayer as shown in Fig.~3(e). In the limit of a thick Ag layer, the magnetoresistance of the trilayer disappears, since most of the applied charge current flows in the Ag layer, which shows no magnetoresistance. Here, $\Delta R/R$ is independent of the Bi layer thickness [see the red solid square in Fig.~3(e)], which supports the interpretation that the Bi/Ag interface is responsible for the observed magnetoresistance.

We develop a simple model that describes the $t_\text{Ag}$ dependence of $\Delta R/R$. For simplicity, we neglect the spin relaxation in the Ag layer due to the long spin diffusion length of Ag, compared with $t_\text{Ag}$. Under this assumption, the ratio of the additional 2D charge current due to the IREE to the applied 2D charge current flowing in the Bi/Ag layer is independent of $t_\text{Ag}$. This allows us to approximate the REMR as the magnetoresistance induced in the Bi/Ag interface layer with the resistivity $R_\text{int}$ and thickness $t_\text{int}$. Here, we define the resistance change of the Bi/Ag interface layer due to the REMR as $\Delta R_\text{int}=R_\text{int}(\beta=0)-R_\text{int}(\beta=90^\circ)$. Thus, with these assumptions, the Bi/Ag/CoFeB trilayer can be modeled as four independent resistors in parallel: the Bi, Ag, CoFeB, and Bi/Ag interface layers. By taking into account the shunting and short-circuit effects~\cite{xu2010anomalous,xu2008scaling,supplementary}, we obtain 
\begin{eqnarray}\label{eq444}
\frac{\Delta R}{R}=\frac{1}{R}\left(\frac{R_{\mathrm{Ag}}}{R_{\mathrm{Ag}}+R_{\mathrm{MRL}}}\right)^{2}\left[\left(\frac{R_{\mathrm{CFB/Bi}}}{R_{\mathrm{CFB/Bi}}+R_{\mathrm{int}}}\right)^{2}\Delta R_{\mathrm{int}}+\left(\frac{R_{\mathrm{int}}}{R_{\mathrm{CFB/Bi}}+R_{\mathrm{int}}}\right)^{2}\Delta R_{\mathrm{CFB/Bi}}\right], \label{thicknessdep}
\end{eqnarray}
where $R_\text{Ag}$ is the $t_\text{Ag}$-dependent resistance of the Ag layer. $R_\text{CFB/Bi}=R(t_\text{Ag}=0)$ is the $t_\text{Ag}$-independent parallel resistance of $R_\text{CFB}$ and $R_\text{Bi}$, where $R_\text{Bi}$ is the resistance of the Bi layer. $\Delta R_\text{CFB/Bi}=R_\text{CFB/Bi}(\beta=0)-R_\text{CFB/Bi}(\beta=90^\circ)$. $R_\text{MRL}$ is the parallel resistance of $R_\text{CFB/Bi}$ and $R_\text{int}$. As shown in Fig.~3(e), the experimentally measured $t_\text{Ag}$ dependence of $\Delta R/R$ is reproduced using Eq.~(\ref{thicknessdep}) by assuming the value of $R_\text{int}/(\Delta R_\text{int}/R_\text{int})$ as shown in Fig.~3(f). Figure~3(f) shows that $R_\text{int}/(\Delta R_\text{int}/R_\text{int})$ decays quickly by increasing $t_\text{Ag}$ with the decay constant of 0.55 nm, since in the thin Ag film range, the Bi/Ag interface is not fully continuous because of imperfect coverage of the surface of the CoFeB layer. By increasing $t_\text{Ag}$, $R_\text{int}/(\Delta R_\text{int}/R_\text{int})$ becomes constant when the continuous Bi/Ag interface is formed.

The magnetoresistance mediated by the diffusive spin current generated at the Bi/Ag interface is further evidenced by measuring the electric resistance by changing the strength of the external magnetic field $\mu_0 {H}$.  As shown in Fig.~4, the resistance of the Bi/Ag/CoFeB trilayer increases with ${\mu_0 H}$ when $ \bf H$ is applied along the $x$ or $z$ direction in spite of the fact that the magnetization of the CoFeB layer is saturated. When $ \bf H$ is applied along the $x$ or $z$ axis, the spin polarization direction $\bm{\sigma}$ of the spin accumulation created by the REE is perpendicular to $ {\bf H}$, since $\bm{\sigma}$ is directed along the $y$ axis. When $ {\bf H}$ and $\bm{\sigma}$ are not collinear, the Hanle effect leads to a spin precession and dephasing of the spin accumulation. Thus, in the trilayer, both the spin accumulation and the additional charge current due to the IREE are partially suppressed by the Hanle spin precession, increasing the resistance with $\mu_0 H$~\cite{PhysRevLett.116.016603}. Figure~4 further demonstrates that the $\mu_0 H$-dependent magnetoresistance is absent when $ \bf H $ is applied parallel to $\bm{\sigma}$, i.e. along the $y$ axis, consistent with the prediction of the Hanle magnetoresistance (HMR).

In summary, we observed a damping-like torque due to the absorption of a diffusive spin current created by the Rashba spin splitting at a Bi/Ag interface. The diffusive spin current couples the spin accumulation at the interface and magnetization in an adjacent ferromagnetic metal, resulting the magnetoresistance through the simultaneous action of the REE and IREE in a Bi/Ag/CoFeB trilayer. The electric resistance of the trilayer has been found to depend on the strength of the external magnetic field, even when the magnetization of the CoFeB layer is saturated, demonstrating that the resistance is affected by the Hanle spin precession in the Ag layer.

\begin{acknowledgments}
This work was supported in part by JSPS KAKENHI Grant Numbers 26790037, 26220604, 26103004, PRESTO-JST ``Innovative nano-electronics through interdisciplinary collaboration among material, device and system layers," the Mitsubishi Foundation, the Mizuho Foundation for the Promotion of Science, the Casio Science Promotion Foundation, and the Murata Science Foundation.
\end{acknowledgments}

\clearpage


\begin{thebibliography}{39}%
\makeatletter
\providecommand \@ifxundefined [1]{%
 \@ifx{#1\undefined}
}%
\providecommand \@ifnum [1]{%
 \ifnum #1\expandafter \@firstoftwo
 \else \expandafter \@secondoftwo
 \fi
}%
\providecommand \@ifx [1]{%
 \ifx #1\expandafter \@firstoftwo
 \else \expandafter \@secondoftwo
 \fi
}%
\providecommand \natexlab [1]{#1}%
\providecommand \enquote  [1]{``#1''}%
\providecommand \bibnamefont  [1]{#1}%
\providecommand \bibfnamefont [1]{#1}%
\providecommand \citenamefont [1]{#1}%
\providecommand \href@noop [0]{\@secondoftwo}%
\providecommand \href [0]{\begingroup \@sanitize@url \@href}%
\providecommand \@href[1]{\@@startlink{#1}\@@href}%
\providecommand \@@href[1]{\endgroup#1\@@endlink}%
\providecommand \@sanitize@url [0]{\catcode `\\12\catcode `\$12\catcode
  `\&12\catcode `\#12\catcode `\^12\catcode `\_12\catcode `\%12\relax}%
\providecommand \@@startlink[1]{}%
\providecommand \@@endlink[0]{}%
\providecommand \url  [0]{\begingroup\@sanitize@url \@url }%
\providecommand \@url [1]{\endgroup\@href {#1}{\urlprefix }}%
\providecommand \urlprefix  [0]{URL }%
\providecommand \Eprint [0]{\href }%
\providecommand \doibase [0]{http://dx.doi.org/}%
\providecommand \selectlanguage [0]{\@gobble}%
\providecommand \bibinfo  [0]{\@secondoftwo}%
\providecommand \bibfield  [0]{\@secondoftwo}%
\providecommand \translation [1]{[#1]}%
\providecommand \BibitemOpen [0]{}%
\providecommand \bibitemStop [0]{}%
\providecommand \bibitemNoStop [0]{.\EOS\space}%
\providecommand \EOS [0]{\spacefactor3000\relax}%
\providecommand \BibitemShut  [1]{\csname bibitem#1\endcsname}%
\let\auto@bib@innerbib\@empty
\bibitem [{\citenamefont {Maekawa}\ and\ \citenamefont
  {Shinjo}(2002)}]{Maekawa2}%
  \BibitemOpen
  \bibinfo {editor} {\bibfnamefont {S.}~\bibnamefont {Maekawa}}\ and\ \bibinfo
  {editor} {\bibfnamefont {T.}~\bibnamefont {Shinjo}},\ eds.,\ \href@noop {}
  {\emph {\bibinfo {title} {Spin Dependent Transport in Magnetic
  Nanostructures}}}\ (\bibinfo  {publisher} {Taylor and Francis},\ \bibinfo
  {address} {New York},\ \bibinfo {year} {2002})\BibitemShut {NoStop}%
\bibitem [{\citenamefont {Nakayama}\ \emph {et~al.}(2013)\citenamefont
  {Nakayama}, \citenamefont {Althammer}, \citenamefont {Chen}, \citenamefont
  {Uchida}, \citenamefont {Kajiwara}, \citenamefont {Kikuchi}, \citenamefont
  {Ohtani}, \citenamefont {Geprags}, \citenamefont {Opel}, \citenamefont
  {Takahashi}, \citenamefont {Gross}, \citenamefont {Bauer}, \citenamefont
  {Goennenwein},\ and\ \citenamefont {Saitoh}}]{Nakayama:206601}%
  \BibitemOpen
  \bibfield  {author} {\bibinfo {author} {\bibfnamefont {H.}~\bibnamefont
  {Nakayama}}, \bibinfo {author} {\bibfnamefont {M.}~\bibnamefont {Althammer}},
  \bibinfo {author} {\bibfnamefont {Y.~T.}\ \bibnamefont {Chen}}, \bibinfo
  {author} {\bibfnamefont {K.}~\bibnamefont {Uchida}}, \bibinfo {author}
  {\bibfnamefont {Y.}~\bibnamefont {Kajiwara}}, \bibinfo {author}
  {\bibfnamefont {D.}~\bibnamefont {Kikuchi}}, \bibinfo {author} {\bibfnamefont
  {T.}~\bibnamefont {Ohtani}}, \bibinfo {author} {\bibfnamefont
  {S.}~\bibnamefont {Geprags}}, \bibinfo {author} {\bibfnamefont
  {M.}~\bibnamefont {Opel}}, \bibinfo {author} {\bibfnamefont {S.}~\bibnamefont
  {Takahashi}}, \bibinfo {author} {\bibfnamefont {R.}~\bibnamefont {Gross}},
  \bibinfo {author} {\bibfnamefont {G.~E.~W.}\ \bibnamefont {Bauer}}, \bibinfo
  {author} {\bibfnamefont {S.~T.~B.}\ \bibnamefont {Goennenwein}}, \ and\
  \bibinfo {author} {\bibfnamefont {E.}~\bibnamefont {Saitoh}},\ }\href@noop {}
  {\bibfield  {journal} {\bibinfo  {journal} {Phys. Rev. Lett.}\ }\textbf
  {\bibinfo {volume} {110}},\ \bibinfo {pages} {206601} (\bibinfo {year}
  {2013})}\BibitemShut {NoStop}%
\bibitem [{\citenamefont {Kim}\ \emph {et~al.}(2016)\citenamefont {Kim},
  \citenamefont {Sheng}, \citenamefont {Takahashi}, \citenamefont {Mitani},\
  and\ \citenamefont {Hayashi}}]{PhysRevLett.116.097201}%
  \BibitemOpen
  \bibfield  {author} {\bibinfo {author} {\bibfnamefont {J.}~\bibnamefont
  {Kim}}, \bibinfo {author} {\bibfnamefont {P.}~\bibnamefont {Sheng}}, \bibinfo
  {author} {\bibfnamefont {S.}~\bibnamefont {Takahashi}}, \bibinfo {author}
  {\bibfnamefont {S.}~\bibnamefont {Mitani}}, \ and\ \bibinfo {author}
  {\bibfnamefont {M.}~\bibnamefont {Hayashi}},\ }\href@noop {} {\bibfield
  {journal} {\bibinfo  {journal} {Phys. Rev. Lett.}\ }\textbf {\bibinfo
  {volume} {116}},\ \bibinfo {pages} {097201} (\bibinfo {year}
  {2016})}\BibitemShut {NoStop}%
\bibitem [{\citenamefont {Chen}\ \emph {et~al.}(2013)\citenamefont {Chen},
  \citenamefont {Takahashi}, \citenamefont {Nakayama}, \citenamefont
  {Althammer}, \citenamefont {Goennenwein}, \citenamefont {Saitoh},\ and\
  \citenamefont {Bauer}}]{chen2013theory}%
  \BibitemOpen
  \bibfield  {author} {\bibinfo {author} {\bibfnamefont {Y.-T.}\ \bibnamefont
  {Chen}}, \bibinfo {author} {\bibfnamefont {S.}~\bibnamefont {Takahashi}},
  \bibinfo {author} {\bibfnamefont {H.}~\bibnamefont {Nakayama}}, \bibinfo
  {author} {\bibfnamefont {M.}~\bibnamefont {Althammer}}, \bibinfo {author}
  {\bibfnamefont {S.~T.}\ \bibnamefont {Goennenwein}}, \bibinfo {author}
  {\bibfnamefont {E.}~\bibnamefont {Saitoh}}, \ and\ \bibinfo {author}
  {\bibfnamefont {G.~E.}\ \bibnamefont {Bauer}},\ }\href@noop {} {\bibfield
  {journal} {\bibinfo  {journal} {Phys. Rev. B}\ }\textbf {\bibinfo {volume}
  {87}},\ \bibinfo {pages} {144411} (\bibinfo {year} {2013})}\BibitemShut
  {NoStop}%
\bibitem [{\citenamefont {Althammer}\ \emph {et~al.}(2013)\citenamefont
  {Althammer}, \citenamefont {Meyer}, \citenamefont {Nakayama}, \citenamefont
  {Schreier}, \citenamefont {Altmannshofer}, \citenamefont {Weiler},
  \citenamefont {Huebl}, \citenamefont {Gepr{\"a}gs}, \citenamefont {Opel},
  \citenamefont {Gross} \emph {et~al.}}]{althammer2013quantitative}%
  \BibitemOpen
  \bibfield  {author} {\bibinfo {author} {\bibfnamefont {M.}~\bibnamefont
  {Althammer}}, \bibinfo {author} {\bibfnamefont {S.}~\bibnamefont {Meyer}},
  \bibinfo {author} {\bibfnamefont {H.}~\bibnamefont {Nakayama}}, \bibinfo
  {author} {\bibfnamefont {M.}~\bibnamefont {Schreier}}, \bibinfo {author}
  {\bibfnamefont {S.}~\bibnamefont {Altmannshofer}}, \bibinfo {author}
  {\bibfnamefont {M.}~\bibnamefont {Weiler}}, \bibinfo {author} {\bibfnamefont
  {H.}~\bibnamefont {Huebl}}, \bibinfo {author} {\bibfnamefont
  {S.}~\bibnamefont {Gepr{\"a}gs}}, \bibinfo {author} {\bibfnamefont
  {M.}~\bibnamefont {Opel}}, \bibinfo {author} {\bibfnamefont {R.}~\bibnamefont
  {Gross}},  \emph {et~al.},\ }\href@noop {} {\bibfield  {journal} {\bibinfo
  {journal} {Phys. Rev. B}\ }\textbf {\bibinfo {volume} {87}},\ \bibinfo
  {pages} {224401} (\bibinfo {year} {2013})}\BibitemShut {NoStop}%
\bibitem [{\citenamefont {Vlietstra}\ \emph {et~al.}(2013)\citenamefont
  {Vlietstra}, \citenamefont {Shan}, \citenamefont {Castel}, \citenamefont {van
  Wees},\ and\ \citenamefont {Youssef}}]{vlietstra2013spin}%
  \BibitemOpen
  \bibfield  {author} {\bibinfo {author} {\bibfnamefont {N.}~\bibnamefont
  {Vlietstra}}, \bibinfo {author} {\bibfnamefont {J.}~\bibnamefont {Shan}},
  \bibinfo {author} {\bibfnamefont {V.}~\bibnamefont {Castel}}, \bibinfo
  {author} {\bibfnamefont {B.}~\bibnamefont {van Wees}}, \ and\ \bibinfo
  {author} {\bibfnamefont {J.~B.}\ \bibnamefont {Youssef}},\ }\href@noop {}
  {\bibfield  {journal} {\bibinfo  {journal} {Phys. Rev. B}\ }\textbf {\bibinfo
  {volume} {87}},\ \bibinfo {pages} {184421} (\bibinfo {year}
  {2013})}\BibitemShut {NoStop}%
\bibitem [{\citenamefont {Hahn}\ \emph {et~al.}(2013)\citenamefont {Hahn},
  \citenamefont {De~Loubens}, \citenamefont {Klein}, \citenamefont {Viret},
  \citenamefont {Naletov},\ and\ \citenamefont
  {Youssef}}]{hahn2013comparative}%
  \BibitemOpen
  \bibfield  {author} {\bibinfo {author} {\bibfnamefont {C.}~\bibnamefont
  {Hahn}}, \bibinfo {author} {\bibfnamefont {G.}~\bibnamefont {De~Loubens}},
  \bibinfo {author} {\bibfnamefont {O.}~\bibnamefont {Klein}}, \bibinfo
  {author} {\bibfnamefont {M.}~\bibnamefont {Viret}}, \bibinfo {author}
  {\bibfnamefont {V.~V.}\ \bibnamefont {Naletov}}, \ and\ \bibinfo {author}
  {\bibfnamefont {J.~B.}\ \bibnamefont {Youssef}},\ }\href@noop {} {\bibfield
  {journal} {\bibinfo  {journal} {Phys. Rev. B}\ }\textbf {\bibinfo {volume}
  {87}},\ \bibinfo {pages} {174417} (\bibinfo {year} {2013})}\BibitemShut
  {NoStop}%
\bibitem [{\citenamefont {Dyakonov}\ and\ \citenamefont
  {Perel}(1971)}]{Dyakonov}%
  \BibitemOpen
  \bibfield  {author} {\bibinfo {author} {\bibfnamefont {M.~I.}\ \bibnamefont
  {Dyakonov}}\ and\ \bibinfo {author} {\bibfnamefont {V.~I.}\ \bibnamefont
  {Perel}},\ }\href@noop {} {\bibfield  {journal} {\bibinfo  {journal} {Phys.
  Lett.}\ }\textbf {\bibinfo {volume} {35A}},\ \bibinfo {pages} {459} (\bibinfo
  {year} {1971})}\BibitemShut {NoStop}%
\bibitem [{\citenamefont {Hirsch}(1999)}]{Hirsch}%
  \BibitemOpen
  \bibfield  {author} {\bibinfo {author} {\bibfnamefont {J.~E.}\ \bibnamefont
  {Hirsch}},\ }\href@noop {} {\bibfield  {journal} {\bibinfo  {journal} {Phys.
  Rev. Lett.}\ }\textbf {\bibinfo {volume} {83}},\ \bibinfo {pages} {1834}
  (\bibinfo {year} {1999})}\BibitemShut {NoStop}%
\bibitem [{\citenamefont {Murakami}\ \emph {et~al.}(2003)\citenamefont
  {Murakami}, \citenamefont {Nagaosa},\ and\ \citenamefont {Zhang}}]{Murakami}%
  \BibitemOpen
  \bibfield  {author} {\bibinfo {author} {\bibfnamefont {S.}~\bibnamefont
  {Murakami}}, \bibinfo {author} {\bibfnamefont {N.}~\bibnamefont {Nagaosa}}, \
  and\ \bibinfo {author} {\bibfnamefont {S.~C.}\ \bibnamefont {Zhang}},\
  }\href@noop {} {\bibfield  {journal} {\bibinfo  {journal} {Science}\ }\textbf
  {\bibinfo {volume} {301}},\ \bibinfo {pages} {1348} (\bibinfo {year}
  {2003})}\BibitemShut {NoStop}%
\bibitem [{\citenamefont {Sinova}\ \emph {et~al.}(2004)\citenamefont {Sinova},
  \citenamefont {Culcer}, \citenamefont {Niu}, \citenamefont {Sinitsyn},
  \citenamefont {Jungwirth},\ and\ \citenamefont {MacDonald}}]{Sinova}%
  \BibitemOpen
  \bibfield  {author} {\bibinfo {author} {\bibfnamefont {J.}~\bibnamefont
  {Sinova}}, \bibinfo {author} {\bibfnamefont {D.}~\bibnamefont {Culcer}},
  \bibinfo {author} {\bibfnamefont {Q.}~\bibnamefont {Niu}}, \bibinfo {author}
  {\bibfnamefont {N.~A.}\ \bibnamefont {Sinitsyn}}, \bibinfo {author}
  {\bibfnamefont {T.}~\bibnamefont {Jungwirth}}, \ and\ \bibinfo {author}
  {\bibfnamefont {A.~H.}\ \bibnamefont {MacDonald}},\ }\href@noop {} {\bibfield
   {journal} {\bibinfo  {journal} {Phys. Rev. Lett.}\ }\textbf {\bibinfo
  {volume} {92}},\ \bibinfo {pages} {126603} (\bibinfo {year}
  {2004})}\BibitemShut {NoStop}%
\bibitem [{\citenamefont {Kato}\ \emph {et~al.}(2004)\citenamefont {Kato},
  \citenamefont {Myers}, \citenamefont {Gossard},\ and\ \citenamefont
  {Awschalom}}]{Kato}%
  \BibitemOpen
  \bibfield  {author} {\bibinfo {author} {\bibfnamefont {Y.~K.}\ \bibnamefont
  {Kato}}, \bibinfo {author} {\bibfnamefont {R.~C.}\ \bibnamefont {Myers}},
  \bibinfo {author} {\bibfnamefont {A.~C.}\ \bibnamefont {Gossard}}, \ and\
  \bibinfo {author} {\bibfnamefont {D.~D.}\ \bibnamefont {Awschalom}},\
  }\href@noop {} {\bibfield  {journal} {\bibinfo  {journal} {Science}\ }\textbf
  {\bibinfo {volume} {306}},\ \bibinfo {pages} {1910} (\bibinfo {year}
  {2004})}\BibitemShut {NoStop}%
\bibitem [{\citenamefont {Wunderlich}\ \emph {et~al.}(2005)\citenamefont
  {Wunderlich}, \citenamefont {Kaestner}, \citenamefont {Sinova},\ and\
  \citenamefont {Jungwirth}}]{Wunderlich}%
  \BibitemOpen
  \bibfield  {author} {\bibinfo {author} {\bibfnamefont {J.}~\bibnamefont
  {Wunderlich}}, \bibinfo {author} {\bibfnamefont {B.}~\bibnamefont
  {Kaestner}}, \bibinfo {author} {\bibfnamefont {J.}~\bibnamefont {Sinova}}, \
  and\ \bibinfo {author} {\bibfnamefont {T.}~\bibnamefont {Jungwirth}},\
  }\href@noop {} {\bibfield  {journal} {\bibinfo  {journal} {Phys. Rev. Lett.}\
  }\textbf {\bibinfo {volume} {94}},\ \bibinfo {pages} {047204} (\bibinfo
  {year} {2005})}\BibitemShut {NoStop}%
\bibitem [{\citenamefont {Sinova}\ \emph {et~al.}(2015)\citenamefont {Sinova},
  \citenamefont {Valenzuela}, \citenamefont {Wunderlich}, \citenamefont
  {Back},\ and\ \citenamefont {Jungwirth}}]{RevModPhys.87.1213}%
  \BibitemOpen
  \bibfield  {author} {\bibinfo {author} {\bibfnamefont {J.}~\bibnamefont
  {Sinova}}, \bibinfo {author} {\bibfnamefont {S.~O.}\ \bibnamefont
  {Valenzuela}}, \bibinfo {author} {\bibfnamefont {J.}~\bibnamefont
  {Wunderlich}}, \bibinfo {author} {\bibfnamefont {C.~H.}\ \bibnamefont
  {Back}}, \ and\ \bibinfo {author} {\bibfnamefont {T.}~\bibnamefont
  {Jungwirth}},\ }\href@noop {} {\bibfield  {journal} {\bibinfo  {journal}
  {Rev. Mod. Phys.}\ }\textbf {\bibinfo {volume} {87}},\ \bibinfo {pages}
  {1213} (\bibinfo {year} {2015})}\BibitemShut {NoStop}%
\bibitem [{\citenamefont {Hoffmann}(2013)}]{6516040}%
  \BibitemOpen
  \bibfield  {author} {\bibinfo {author} {\bibfnamefont {A.}~\bibnamefont
  {Hoffmann}},\ }\href@noop {} {\bibfield  {journal} {\bibinfo  {journal} {IEEE
  Trans. Magn.}\ }\textbf {\bibinfo {volume} {49}},\ \bibinfo {pages} {5172}
  (\bibinfo {year} {2013})}\BibitemShut {NoStop}%
\bibitem [{\citenamefont {Valenzuela}\ and\ \citenamefont
  {Tinkham}(2006)}]{Valenzuela}%
  \BibitemOpen
  \bibfield  {author} {\bibinfo {author} {\bibfnamefont {S.~O.}\ \bibnamefont
  {Valenzuela}}\ and\ \bibinfo {author} {\bibfnamefont {M.}~\bibnamefont
  {Tinkham}},\ }\href@noop {} {\bibfield  {journal} {\bibinfo  {journal}
  {Nature}\ }\textbf {\bibinfo {volume} {442}},\ \bibinfo {pages} {176}
  (\bibinfo {year} {2006})}\BibitemShut {NoStop}%
\bibitem [{\citenamefont {Kimura}\ \emph {et~al.}(2007)\citenamefont {Kimura},
  \citenamefont {Otani}, \citenamefont {Sato}, \citenamefont {Takahashi},\ and\
  \citenamefont {Maekawa}}]{KimuraPRL}%
  \BibitemOpen
  \bibfield  {author} {\bibinfo {author} {\bibfnamefont {T.}~\bibnamefont
  {Kimura}}, \bibinfo {author} {\bibfnamefont {Y.}~\bibnamefont {Otani}},
  \bibinfo {author} {\bibfnamefont {T.}~\bibnamefont {Sato}}, \bibinfo {author}
  {\bibfnamefont {S.}~\bibnamefont {Takahashi}}, \ and\ \bibinfo {author}
  {\bibfnamefont {S.}~\bibnamefont {Maekawa}},\ }\href@noop {} {\bibfield
  {journal} {\bibinfo  {journal} {Phys. Rev. Lett.}\ }\textbf {\bibinfo
  {volume} {98}},\ \bibinfo {eid} {156601} (\bibinfo {year}
  {2007})}\BibitemShut {NoStop}%
\bibitem [{\citenamefont {Saitoh}\ \emph {et~al.}(2006)\citenamefont {Saitoh},
  \citenamefont {Ueda}, \citenamefont {Miyajima},\ and\ \citenamefont
  {Tatara}}]{Saitoh}%
  \BibitemOpen
  \bibfield  {author} {\bibinfo {author} {\bibfnamefont {E.}~\bibnamefont
  {Saitoh}}, \bibinfo {author} {\bibfnamefont {M.}~\bibnamefont {Ueda}},
  \bibinfo {author} {\bibfnamefont {H.}~\bibnamefont {Miyajima}}, \ and\
  \bibinfo {author} {\bibfnamefont {G.}~\bibnamefont {Tatara}},\ }\href@noop {}
  {\bibfield  {journal} {\bibinfo  {journal} {Appl. Phys. Lett.}\ }\textbf
  {\bibinfo {volume} {88}},\ \bibinfo {pages} {182509} (\bibinfo {year}
  {2006})}\BibitemShut {NoStop}%
\bibitem [{\citenamefont {Edelstein}(1990)}]{edelstein1990spin}%
  \BibitemOpen
  \bibfield  {author} {\bibinfo {author} {\bibfnamefont {V.~M.}\ \bibnamefont
  {Edelstein}},\ }\href@noop {} {\bibfield  {journal} {\bibinfo  {journal}
  {Solid State Commun.}\ }\textbf {\bibinfo {volume} {73}},\ \bibinfo {pages}
  {233} (\bibinfo {year} {1990})}\BibitemShut {NoStop}%
\bibitem [{\citenamefont {Rojas-Sanchez}\ \emph {et~al.}(2013)\citenamefont
  {Rojas-Sanchez}, \citenamefont {Vila}, \citenamefont {Desfonds},
  \citenamefont {Gambarelli}, \citenamefont {Attane}, \citenamefont {Teresa},
  \citenamefont {Magen},\ and\ \citenamefont {Fert}}]{Sanchez}%
  \BibitemOpen
  \bibfield  {author} {\bibinfo {author} {\bibfnamefont {J.~C.}\ \bibnamefont
  {Rojas-Sanchez}}, \bibinfo {author} {\bibfnamefont {L.}~\bibnamefont {Vila}},
  \bibinfo {author} {\bibfnamefont {G.}~\bibnamefont {Desfonds}}, \bibinfo
  {author} {\bibfnamefont {S.}~\bibnamefont {Gambarelli}}, \bibinfo {author}
  {\bibfnamefont {J.~P.}\ \bibnamefont {Attane}}, \bibinfo {author}
  {\bibfnamefont {J.~M.~D.}\ \bibnamefont {Teresa}}, \bibinfo {author}
  {\bibfnamefont {C.}~\bibnamefont {Magen}}, \ and\ \bibinfo {author}
  {\bibfnamefont {A.}~\bibnamefont {Fert}},\ }\href@noop {} {\bibfield
  {journal} {\bibinfo  {journal} {Nat. Commun.}\ }\textbf {\bibinfo {volume}
  {4}},\ \bibinfo {pages} {2944} (\bibinfo {year} {2013})}\BibitemShut
  {NoStop}%
\bibitem [{\citenamefont {Sangiao}\ \emph {et~al.}(2015)\citenamefont
  {Sangiao}, \citenamefont {Teresa}, \citenamefont {Morellon}, \citenamefont
  {Lucas}, \citenamefont {Martinez-Valarte},\ and\ \citenamefont
  {Viret}}]{Sangiao}%
  \BibitemOpen
  \bibfield  {author} {\bibinfo {author} {\bibfnamefont {S.}~\bibnamefont
  {Sangiao}}, \bibinfo {author} {\bibfnamefont {J.~M.~D.}\ \bibnamefont
  {Teresa}}, \bibinfo {author} {\bibfnamefont {L.}~\bibnamefont {Morellon}},
  \bibinfo {author} {\bibfnamefont {I.}~\bibnamefont {Lucas}}, \bibinfo
  {author} {\bibfnamefont {M.~C.}\ \bibnamefont {Martinez-Valarte}}, \ and\
  \bibinfo {author} {\bibfnamefont {M.}~\bibnamefont {Viret}},\ }\href@noop {}
  {\bibfield  {journal} {\bibinfo  {journal} {Appl. Phys. Lett.}\ }\textbf
  {\bibinfo {volume} {106}},\ \bibinfo {pages} {172403} (\bibinfo {year}
  {2015})}\BibitemShut {NoStop}%
\bibitem [{\citenamefont {Nomura}\ \emph {et~al.}(2015)\citenamefont {Nomura},
  \citenamefont {Tashiro}, \citenamefont {Nakayama},\ and\ \citenamefont
  {Ando}}]{Nomura:212403}%
  \BibitemOpen
  \bibfield  {author} {\bibinfo {author} {\bibfnamefont {A.}~\bibnamefont
  {Nomura}}, \bibinfo {author} {\bibfnamefont {T.}~\bibnamefont {Tashiro}},
  \bibinfo {author} {\bibfnamefont {H.}~\bibnamefont {Nakayama}}, \ and\
  \bibinfo {author} {\bibfnamefont {K.}~\bibnamefont {Ando}},\ }\href@noop {}
  {\bibfield  {journal} {\bibinfo  {journal} {Appl. Phys. Lett.}\ }\textbf
  {\bibinfo {volume} {106}},\ \bibinfo {pages} {212403} (\bibinfo {year}
  {2015})}\BibitemShut {NoStop}%
\bibitem [{\citenamefont {Jungfleisch}\ \emph {et~al.}(2016)\citenamefont
  {Jungfleisch}, \citenamefont {Zhang}, \citenamefont {Sklenar}, \citenamefont
  {Jiang}, \citenamefont {Pearson}, \citenamefont {Ketterson},\ and\
  \citenamefont {Hoffmann}}]{PhysRevB.93.2244192}%
  \BibitemOpen
  \bibfield  {author} {\bibinfo {author} {\bibfnamefont {M.~B.}\ \bibnamefont
  {Jungfleisch}}, \bibinfo {author} {\bibfnamefont {W.}~\bibnamefont {Zhang}},
  \bibinfo {author} {\bibfnamefont {J.}~\bibnamefont {Sklenar}}, \bibinfo
  {author} {\bibfnamefont {W.}~\bibnamefont {Jiang}}, \bibinfo {author}
  {\bibfnamefont {J.~E.}\ \bibnamefont {Pearson}}, \bibinfo {author}
  {\bibfnamefont {J.~B.}\ \bibnamefont {Ketterson}}, \ and\ \bibinfo {author}
  {\bibfnamefont {A.}~\bibnamefont {Hoffmann}},\ }\href {\doibase
  10.1103/PhysRevB.93.224419} {\bibfield  {journal} {\bibinfo  {journal} {Phys.
  Rev. B}\ }\textbf {\bibinfo {volume} {93}},\ \bibinfo {pages} {224419}
  (\bibinfo {year} {2016})}\BibitemShut {NoStop}%
\bibitem [{\citenamefont {Isasa}\ \emph {et~al.}(2016)\citenamefont {Isasa},
  \citenamefont {Martinez-Valarte}, \citenamefont {Villamor}, \citenamefont
  {Magen}, \citenamefont {Morellon}, \citenamefont {Teresa}, \citenamefont
  {Ibarra}, \citenamefont {Vignale}, \citenamefont {Chulkov}, \citenamefont
  {Krasovskii}, \citenamefont {Hueso},\ and\ \citenamefont
  {Casanova}}]{Isasa:014420}%
  \BibitemOpen
  \bibfield  {author} {\bibinfo {author} {\bibfnamefont {M.}~\bibnamefont
  {Isasa}}, \bibinfo {author} {\bibfnamefont {M.~C.}\ \bibnamefont
  {Martinez-Valarte}}, \bibinfo {author} {\bibfnamefont {E.}~\bibnamefont
  {Villamor}}, \bibinfo {author} {\bibfnamefont {C.}~\bibnamefont {Magen}},
  \bibinfo {author} {\bibfnamefont {L.}~\bibnamefont {Morellon}}, \bibinfo
  {author} {\bibfnamefont {J.~M.~D.}\ \bibnamefont {Teresa}}, \bibinfo {author}
  {\bibfnamefont {M.~R.}\ \bibnamefont {Ibarra}}, \bibinfo {author}
  {\bibfnamefont {G.}~\bibnamefont {Vignale}}, \bibinfo {author} {\bibfnamefont
  {E.~V.}\ \bibnamefont {Chulkov}}, \bibinfo {author} {\bibfnamefont {E.~E.}\
  \bibnamefont {Krasovskii}}, \bibinfo {author} {\bibfnamefont {L.~E.}\
  \bibnamefont {Hueso}}, \ and\ \bibinfo {author} {\bibfnamefont
  {F.}~\bibnamefont {Casanova}},\ }\href@noop {} {\bibfield  {journal}
  {\bibinfo  {journal} {Phys. Rev. B}\ }\textbf {\bibinfo {volume} {93}},\
  \bibinfo {pages} {014420} (\bibinfo {year} {2016})}\BibitemShut {NoStop}%
\bibitem [{\citenamefont {Zhang}\ \emph {et~al.}(2016)\citenamefont {Zhang},
  \citenamefont {Yamamoto}, \citenamefont {Gu}, \citenamefont {Li},
  \citenamefont {Maekawa}, \citenamefont {Fukaya},\ and\ \citenamefont
  {Kawasuso}}]{Zhang:014420}%
  \BibitemOpen
  \bibfield  {author} {\bibinfo {author} {\bibfnamefont {H.~J.}\ \bibnamefont
  {Zhang}}, \bibinfo {author} {\bibfnamefont {S.}~\bibnamefont {Yamamoto}},
  \bibinfo {author} {\bibfnamefont {B.}~\bibnamefont {Gu}}, \bibinfo {author}
  {\bibfnamefont {H.}~\bibnamefont {Li}}, \bibinfo {author} {\bibfnamefont
  {M.}~\bibnamefont {Maekawa}}, \bibinfo {author} {\bibfnamefont
  {Y.}~\bibnamefont {Fukaya}}, \ and\ \bibinfo {author} {\bibfnamefont
  {A.}~\bibnamefont {Kawasuso}},\ }\href@noop {} {\bibfield  {journal}
  {\bibinfo  {journal} {Phys. Rev. Lett.}\ }\textbf {\bibinfo {volume} {114}},\
  \bibinfo {pages} {166602} (\bibinfo {year} {2016})}\BibitemShut {NoStop}%
\bibitem [{\citenamefont {Kim}\ \emph {et~al.}(2013)\citenamefont {Kim},
  \citenamefont {Sinha}, \citenamefont {Hayashi}, \citenamefont {Yamanouchi},
  \citenamefont {Fukami}, \citenamefont {Suzuki}, \citenamefont {Mitani},\ and\
  \citenamefont {Ohno}}]{Kim:240}%
  \BibitemOpen
  \bibfield  {author} {\bibinfo {author} {\bibfnamefont {J.}~\bibnamefont
  {Kim}}, \bibinfo {author} {\bibfnamefont {J.}~\bibnamefont {Sinha}}, \bibinfo
  {author} {\bibfnamefont {M.}~\bibnamefont {Hayashi}}, \bibinfo {author}
  {\bibfnamefont {M.}~\bibnamefont {Yamanouchi}}, \bibinfo {author}
  {\bibfnamefont {S.}~\bibnamefont {Fukami}}, \bibinfo {author} {\bibfnamefont
  {T.}~\bibnamefont {Suzuki}}, \bibinfo {author} {\bibfnamefont
  {S.}~\bibnamefont {Mitani}}, \ and\ \bibinfo {author} {\bibfnamefont
  {H.}~\bibnamefont {Ohno}},\ }\href@noop {} {\bibfield  {journal} {\bibinfo
  {journal} {Nat. Mater.}\ }\textbf {\bibinfo {volume} {12}},\ \bibinfo {pages}
  {240} (\bibinfo {year} {2013})}\BibitemShut {NoStop}%
\bibitem [{\citenamefont {Hayashi}\ \emph {et~al.}(2014)\citenamefont
  {Hayashi}, \citenamefont {Kim}, \citenamefont {Yamanouchi},\ and\
  \citenamefont {Ohno}}]{Hayashi:144425}%
  \BibitemOpen
  \bibfield  {author} {\bibinfo {author} {\bibfnamefont {M.}~\bibnamefont
  {Hayashi}}, \bibinfo {author} {\bibfnamefont {J.}~\bibnamefont {Kim}},
  \bibinfo {author} {\bibfnamefont {M.}~\bibnamefont {Yamanouchi}}, \ and\
  \bibinfo {author} {\bibfnamefont {H.}~\bibnamefont {Ohno}},\ }\href@noop {}
  {\bibfield  {journal} {\bibinfo  {journal} {Phys. Rev. B}\ }\textbf {\bibinfo
  {volume} {89}},\ \bibinfo {pages} {144425} (\bibinfo {year}
  {2014})}\BibitemShut {NoStop}%
\bibitem [{\citenamefont {Yang}\ \emph {et~al.}(2014)\citenamefont {Yang},
  \citenamefont {Kohda}, \citenamefont {Seki}, \citenamefont {Takanashi},\ and\
  \citenamefont {Nitta}}]{Yang:04EM06}%
  \BibitemOpen
  \bibfield  {author} {\bibinfo {author} {\bibfnamefont {T.}~\bibnamefont
  {Yang}}, \bibinfo {author} {\bibfnamefont {M.}~\bibnamefont {Kohda}},
  \bibinfo {author} {\bibfnamefont {T.}~\bibnamefont {Seki}}, \bibinfo {author}
  {\bibfnamefont {K.}~\bibnamefont {Takanashi}}, \ and\ \bibinfo {author}
  {\bibfnamefont {J.}~\bibnamefont {Nitta}},\ }\href@noop {} {\bibfield
  {journal} {\bibinfo  {journal} {Jpn. J. Appl. Phys.}\ }\textbf {\bibinfo
  {volume} {53}},\ \bibinfo {pages} {04EM06} (\bibinfo {year}
  {2014})}\BibitemShut {NoStop}%
\bibitem [{\citenamefont {Emori}\ \emph {et~al.}(2013)\citenamefont {Emori},
  \citenamefont {Bauer}, \citenamefont {Ahn}, \citenamefont {Martinez},\ and\
  \citenamefont {Beach}}]{Emori:611}%
  \BibitemOpen
  \bibfield  {author} {\bibinfo {author} {\bibfnamefont {S.}~\bibnamefont
  {Emori}}, \bibinfo {author} {\bibfnamefont {U.}~\bibnamefont {Bauer}},
  \bibinfo {author} {\bibfnamefont {S.-M.}\ \bibnamefont {Ahn}}, \bibinfo
  {author} {\bibfnamefont {E.}~\bibnamefont {Martinez}}, \ and\ \bibinfo
  {author} {\bibfnamefont {G.~S.~D.}\ \bibnamefont {Beach}},\ }\href@noop {}
  {\bibfield  {journal} {\bibinfo  {journal} {Nat. Mater.}\ }\textbf {\bibinfo
  {volume} {12}},\ \bibinfo {pages} {611} (\bibinfo {year} {2013})}\BibitemShut
  {NoStop}%
\bibitem [{\citenamefont {Garello}\ \emph {et~al.}(2013)\citenamefont
  {Garello}, \citenamefont {Miron}, \citenamefont {Avci}, \citenamefont
  {Freimuth}, \citenamefont {Makrousov}, \citenamefont {Blugel}, \citenamefont
  {Auffret}, \citenamefont {Boulle}, \citenamefont {Gaudin},\ and\
  \citenamefont {Gambardella}}]{Garello:587}%
  \BibitemOpen
  \bibfield  {author} {\bibinfo {author} {\bibfnamefont {K.}~\bibnamefont
  {Garello}}, \bibinfo {author} {\bibfnamefont {I.~M.}\ \bibnamefont {Miron}},
  \bibinfo {author} {\bibfnamefont {C.~O.}\ \bibnamefont {Avci}}, \bibinfo
  {author} {\bibfnamefont {F.}~\bibnamefont {Freimuth}}, \bibinfo {author}
  {\bibfnamefont {Y.}~\bibnamefont {Makrousov}}, \bibinfo {author}
  {\bibfnamefont {S.}~\bibnamefont {Blugel}}, \bibinfo {author} {\bibfnamefont
  {S.}~\bibnamefont {Auffret}}, \bibinfo {author} {\bibfnamefont
  {O.}~\bibnamefont {Boulle}}, \bibinfo {author} {\bibfnamefont
  {G.}~\bibnamefont {Gaudin}}, \ and\ \bibinfo {author} {\bibfnamefont
  {P.}~\bibnamefont {Gambardella}},\ }\href@noop {} {\bibfield  {journal}
  {\bibinfo  {journal} {Nat. Nanotechnol.}\ }\textbf {\bibinfo {volume} {8}},\
  \bibinfo {pages} {587} (\bibinfo {year} {2013})}\BibitemShut {NoStop}%
\bibitem [{\citenamefont {Haney}\ \emph {et~al.}(2013)\citenamefont {Haney},
  \citenamefont {Lee}, \citenamefont {Lee}, \citenamefont {Manchon},\ and\
  \citenamefont {Stiles}}]{Haney:174411}%
  \BibitemOpen
  \bibfield  {author} {\bibinfo {author} {\bibfnamefont {P.~M.}\ \bibnamefont
  {Haney}}, \bibinfo {author} {\bibfnamefont {H.-W.}\ \bibnamefont {Lee}},
  \bibinfo {author} {\bibfnamefont {K.-J.}\ \bibnamefont {Lee}}, \bibinfo
  {author} {\bibfnamefont {A.}~\bibnamefont {Manchon}}, \ and\ \bibinfo
  {author} {\bibfnamefont {M.~D.}\ \bibnamefont {Stiles}},\ }\href@noop {}
  {\bibfield  {journal} {\bibinfo  {journal} {Phys. Rev. B}\ }\textbf {\bibinfo
  {volume} {87}},\ \bibinfo {pages} {174411} (\bibinfo {year}
  {2013})}\BibitemShut {NoStop}%
\bibitem [{\citenamefont {{An}}\ \emph {et~al.}()\citenamefont {{An}},
  \citenamefont {{Nakayama}}, \citenamefont {{Kanno}}, \citenamefont
  {{Nomura}}, \citenamefont {{Haku}},\ and\ \citenamefont {{Ando}}}]{An:16XX}%
  \BibitemOpen
  \bibfield  {author} {\bibinfo {author} {\bibfnamefont {H.}~\bibnamefont
  {{An}}}, \bibinfo {author} {\bibfnamefont {H.}~\bibnamefont {{Nakayama}}},
  \bibinfo {author} {\bibfnamefont {Y.}~\bibnamefont {{Kanno}}}, \bibinfo
  {author} {\bibfnamefont {A.}~\bibnamefont {{Nomura}}}, \bibinfo {author}
  {\bibfnamefont {S.}~\bibnamefont {{Haku}}}, \ and\ \bibinfo {author}
  {\bibfnamefont {K.}~\bibnamefont {{Ando}}},\ }\href@noop {} {\ }\Eprint
  {http://arxiv.org/abs/1606.05986} {arXiv:1606.05986} \BibitemShut {NoStop}%
\bibitem [{sup()}]{supplementary}%
  \BibitemOpen
  \href@noop {} {}\bibinfo {note} {See Supplemental Material [URL], which
  includes Refs.~[34-44].}\BibitemShut {Stop}%
  \bibitem [{\citenamefont {Yang}\ \emph {et~al.}(2016)\citenamefont {Yang},
  \citenamefont {Cai}, \citenamefont {Ju}, \citenamefont {Edmonds},
  \citenamefont {Yang}, \citenamefont {Liu}, \citenamefont {Li}, \citenamefont
  {Zhang}, \citenamefont {Sheng}, \citenamefont {Wang}, \citenamefont {Ji},\
  and\ \citenamefont {Wang}}]{Yang:20778}%
  \BibitemOpen
  \bibfield  {author} {\bibinfo {author} {\bibfnamefont {M.}~\bibnamefont
  {Yang}}, \bibinfo {author} {\bibfnamefont {K.}~\bibnamefont {Cai}}, \bibinfo
  {author} {\bibfnamefont {H.}~\bibnamefont {Ju}}, \bibinfo {author}
  {\bibfnamefont {K.~W.}\ \bibnamefont {Edmonds}}, \bibinfo {author}
  {\bibfnamefont {G.}~\bibnamefont {Yang}}, \bibinfo {author} {\bibfnamefont
  {S.}~\bibnamefont {Liu}}, \bibinfo {author} {\bibfnamefont {B.}~\bibnamefont
  {Li}}, \bibinfo {author} {\bibfnamefont {B.}~\bibnamefont {Zhang}}, \bibinfo
  {author} {\bibfnamefont {Y.}~\bibnamefont {Sheng}}, \bibinfo {author}
  {\bibfnamefont {S.}~\bibnamefont {Wang}}, \bibinfo {author} {\bibfnamefont
  {Y.}~\bibnamefont {Ji}}, \ and\ \bibinfo {author} {\bibfnamefont
  {K.}~\bibnamefont {Wang}},\ }\href@noop {} {\bibfield  {journal} {\bibinfo
  {journal} {Sci. Rep.}\ }\textbf {\bibinfo {volume} {6}},\ \bibinfo {pages}
  {20778} (\bibinfo {year} {2016})}\BibitemShut {NoStop}%
  \bibitem [{\citenamefont {Miron}\ \emph {et~al.}(2011)\citenamefont {Miron},
  \citenamefont {Garello}, \citenamefont {Gaudin}, \citenamefont {Zermatten},
  \citenamefont {Costache}, \citenamefont {Auffret}, \citenamefont {Bandiera},
  \citenamefont {Rodmacq}, \citenamefont {Schuhl},\ and\ \citenamefont
  {Gambardella}}]{Miron:189}%
  \BibitemOpen
  \bibfield  {author} {\bibinfo {author} {\bibfnamefont {I.~M.}\ \bibnamefont
  {Miron}}, \bibinfo {author} {\bibfnamefont {K.}~\bibnamefont {Garello}},
  \bibinfo {author} {\bibfnamefont {G.}~\bibnamefont {Gaudin}}, \bibinfo
  {author} {\bibfnamefont {P.-J.}\ \bibnamefont {Zermatten}}, \bibinfo {author}
  {\bibfnamefont {M.~V.}\ \bibnamefont {Costache}}, \bibinfo {author}
  {\bibfnamefont {S.}~\bibnamefont {Auffret}}, \bibinfo {author} {\bibfnamefont
  {S.}~\bibnamefont {Bandiera}}, \bibinfo {author} {\bibfnamefont
  {B.}~\bibnamefont {Rodmacq}}, \bibinfo {author} {\bibfnamefont
  {A.}~\bibnamefont {Schuhl}}, \ and\ \bibinfo {author} {\bibfnamefont
  {P.}~\bibnamefont {Gambardella}},\ }\href@noop {} {\bibfield  {journal}
  {\bibinfo  {journal} {Nature}\ }\textbf {\bibinfo {volume} {476}},\ \bibinfo
  {pages} {189} (\bibinfo {year} {2011})}\BibitemShut {NoStop}%
  \bibitem [{\citenamefont {Ghosh}\ \emph {et~al.}(2012)\citenamefont {Ghosh},
  \citenamefont {Auffret}, \citenamefont {Ebels},\ and\ \citenamefont
  {Bailey}}]{Ghosh:127202}%
  \BibitemOpen
  \bibfield  {author} {\bibinfo {author} {\bibfnamefont {A.}~\bibnamefont
  {Ghosh}}, \bibinfo {author} {\bibfnamefont {S.}~\bibnamefont {Auffret}},
  \bibinfo {author} {\bibfnamefont {U.}~\bibnamefont {Ebels}}, \ and\ \bibinfo
  {author} {\bibfnamefont {W.~E.}\ \bibnamefont {Bailey}},\ }\href@noop {}
  {\bibfield  {journal} {\bibinfo  {journal} {Phys. Rev. Lett.}\ }\textbf
  {\bibinfo {volume} {109}},\ \bibinfo {pages} {127202} (\bibinfo {year}
  {2012})}\BibitemShut {NoStop}%
\bibitem [{\citenamefont {Zhang}\ \emph {et~al.}(2015)\citenamefont {Zhang},
  \citenamefont {Han}, \citenamefont {Jiang}, \citenamefont {Yang},\ and\
  \citenamefont {Parkin}}]{Zhang:496}%
  \BibitemOpen
  \bibfield  {author} {\bibinfo {author} {\bibfnamefont {W.}~\bibnamefont
  {Zhang}}, \bibinfo {author} {\bibfnamefont {W.}~\bibnamefont {Han}}, \bibinfo
  {author} {\bibfnamefont {X.}~\bibnamefont {Jiang}}, \bibinfo {author}
  {\bibfnamefont {S.-H.}\ \bibnamefont {Yang}}, \ and\ \bibinfo {author}
  {\bibfnamefont {S.~S.~P.}\ \bibnamefont {Parkin}},\ }\href@noop {} {\bibfield
   {journal} {\bibinfo  {journal} {Nat. Phys.}\ }\textbf {\bibinfo {volume}
  {11}},\ \bibinfo {pages} {496} (\bibinfo {year} {2015})}\BibitemShut
  {NoStop}%
 \bibitem [{\citenamefont {Gambardella}\ and\ \citenamefont
  {Miron}(2011)}]{Gambardella:3175}%
  \BibitemOpen
  \bibfield  {author} {\bibinfo {author} {\bibfnamefont {P.}~\bibnamefont
  {Gambardella}}\ and\ \bibinfo {author} {\bibfnamefont {I.~M.}\ \bibnamefont
  {Miron}},\ }\href@noop {} {\bibfield  {journal} {\bibinfo  {journal} {Philos.
  Trans. Math. Phys. Eng. Sci.}\ }\textbf {\bibinfo {volume} {369}},\ \bibinfo
  {pages} {3175} (\bibinfo {year} {2011})}\BibitemShut {NoStop}%
  \bibitem [{\citenamefont {Godfrey}\ and\ \citenamefont
  {Johnson}(2006)}]{Godfrey:136601}%
  \BibitemOpen
  \bibfield  {author} {\bibinfo {author} {\bibfnamefont {R.}~\bibnamefont
  {Godfrey}}\ and\ \bibinfo {author} {\bibfnamefont {M.}~\bibnamefont
  {Johnson}},\ }\href@noop {} {\bibfield  {journal} {\bibinfo  {journal} {Phys.
  Rev. Lett.}\ }\textbf {\bibinfo {volume} {96}},\ \bibinfo {pages} {136601}
  (\bibinfo {year} {2006})}\BibitemShut {NoStop}%
  \bibitem [{\citenamefont {Sondheimer}(1952)}]{Sondheimer:1}%
  \BibitemOpen
  \bibfield  {author} {\bibinfo {author} {\bibfnamefont {E.~H.}\ \bibnamefont
  {Sondheimer}},\ }\href@noop {} {\bibfield  {journal} {\bibinfo  {journal}
  {Adv. Phys.}\ }\textbf {\bibinfo {volume} {1}},\ \bibinfo {pages} {1}
  (\bibinfo {year} {1952})}\BibitemShut {NoStop}%
\bibitem [{\citenamefont {Fuchs}(1938)}]{Fuchs:100}%
  \BibitemOpen
  \bibfield  {author} {\bibinfo {author} {\bibfnamefont {K.}~\bibnamefont
  {Fuchs}},\ }\href@noop {} {\bibfield  {journal} {\bibinfo  {journal} {Math.
  Proc. Cambridge Philos. Soc.}\ }\textbf {\bibinfo {volume} {34}},\ \bibinfo
  {pages} {100} (\bibinfo {year} {1938})}\BibitemShut {NoStop}%
\bibitem [{\citenamefont {Shiomi}\ and\ \citenamefont
  {Saitoh}(2016)}]{Shiomi:22085}%
  \BibitemOpen
  \bibfield  {author} {\bibinfo {author} {\bibfnamefont {Y.}~\bibnamefont
  {Shiomi}}\ and\ \bibinfo {author} {\bibfnamefont {E.}~\bibnamefont
  {Saitoh}},\ }\href@noop {} {\bibfield  {journal} {\bibinfo  {journal} {Sci.
  Rep.}\ }\textbf {\bibinfo {volume} {6}},\ \bibinfo {pages} {22085} (\bibinfo
  {year} {2016})}\BibitemShut {NoStop}%
\bibitem [{\citenamefont {Cho}\ \emph {et~al.}(2015)\citenamefont {Cho},
  \citenamefont {Baek}, \citenamefont {Lee}, \citenamefont {Jo},\ and\
  \citenamefont {Park}}]{cho2015large}%
  \BibitemOpen
  \bibfield  {author} {\bibinfo {author} {\bibfnamefont {S.}~\bibnamefont
  {Cho}}, \bibinfo {author} {\bibfnamefont {S.-h.~C.}\ \bibnamefont {Baek}},
  \bibinfo {author} {\bibfnamefont {K.-D.}\ \bibnamefont {Lee}}, \bibinfo
  {author} {\bibfnamefont {Y.}~\bibnamefont {Jo}}, \ and\ \bibinfo {author}
  {\bibfnamefont {B.-G.}\ \bibnamefont {Park}},\ }\href@noop {} {\bibfield
  {journal} {\bibinfo  {journal} {Sci. Rep.}\ }\textbf {\bibinfo {volume} {5}}
  (\bibinfo {year} {2015})}\BibitemShut {NoStop}%
\bibitem [{\citenamefont {Hou}\ \emph {et~al.}(2012)\citenamefont {Hou},
  \citenamefont {Qiu}, \citenamefont {Harii}, \citenamefont {Kajiwara},
  \citenamefont {Uchida}, \citenamefont {Fujikawa}, \citenamefont {Nakayama},
  \citenamefont {Yoshino}, \citenamefont {An}, \citenamefont {Ando},
  \citenamefont {Jin},\ and\ \citenamefont {Saitoh}}]{Hou}%
  \BibitemOpen
  \bibfield  {author} {\bibinfo {author} {\bibfnamefont {D.}~\bibnamefont
  {Hou}}, \bibinfo {author} {\bibfnamefont {Z.}~\bibnamefont {Qiu}}, \bibinfo
  {author} {\bibfnamefont {K.}~\bibnamefont {Harii}}, \bibinfo {author}
  {\bibfnamefont {Y.}~\bibnamefont {Kajiwara}}, \bibinfo {author}
  {\bibfnamefont {K.}~\bibnamefont {Uchida}}, \bibinfo {author} {\bibfnamefont
  {Y.}~\bibnamefont {Fujikawa}}, \bibinfo {author} {\bibfnamefont
  {H.}~\bibnamefont {Nakayama}}, \bibinfo {author} {\bibfnamefont
  {T.}~\bibnamefont {Yoshino}}, \bibinfo {author} {\bibfnamefont
  {T.}~\bibnamefont {An}}, \bibinfo {author} {\bibfnamefont {K.}~\bibnamefont
  {Ando}}, \bibinfo {author} {\bibfnamefont {X.}~\bibnamefont {Jin}}, \ and\
  \bibinfo {author} {\bibfnamefont {E.}~\bibnamefont {Saitoh}},\ }\href@noop {}
  {\bibfield  {journal} {\bibinfo  {journal} {Appl. Phys. Lett.}\ }\textbf
  {\bibinfo {volume} {101}},\ \bibinfo {pages} {042403} (\bibinfo {year}
  {2012})}\BibitemShut {NoStop}%
\bibitem [{\citenamefont {Wang}\ \emph {et~al.}(2014)\citenamefont {Wang},
  \citenamefont {Du}, \citenamefont {Pu}, \citenamefont {Adur}, \citenamefont
  {Hammel},\ and\ \citenamefont {Yang}}]{Wang:197201}%
  \BibitemOpen
  \bibfield  {author} {\bibinfo {author} {\bibfnamefont {H.~L.}\ \bibnamefont
  {Wang}}, \bibinfo {author} {\bibfnamefont {C.~H.}\ \bibnamefont {Du}},
  \bibinfo {author} {\bibfnamefont {Y.}~\bibnamefont {Pu}}, \bibinfo {author}
  {\bibfnamefont {R.}~\bibnamefont {Adur}}, \bibinfo {author} {\bibfnamefont
  {P.~C.}\ \bibnamefont {Hammel}}, \ and\ \bibinfo {author} {\bibfnamefont
  {F.~Y.}\ \bibnamefont {Yang}},\ }\href@noop {} {\bibfield  {journal}
  {\bibinfo  {journal} {Phys. Rev. Lett.}\ }\textbf {\bibinfo {volume} {112}},\
  \bibinfo {pages} {197201} (\bibinfo {year} {2014})}\BibitemShut {NoStop}%
\bibitem [{\citenamefont {Rojas-Sanchez}\ \emph {et~al.}(2014)\citenamefont
  {Rojas-Sanchez}, \citenamefont {Reyren}, \citenamefont {Laczkowski},
  \citenamefont {Savero}, \citenamefont {Attane}, \citenamefont {Deranlot},
  \citenamefont {Jamet}, \citenamefont {George}, \citenamefont {Vila},\ and\
  \citenamefont {Jaffres}}]{Sanchez:106602}%
  \BibitemOpen
  \bibfield  {author} {\bibinfo {author} {\bibfnamefont {J.-C.}\ \bibnamefont
  {Rojas-Sanchez}}, \bibinfo {author} {\bibfnamefont {N.}~\bibnamefont
  {Reyren}}, \bibinfo {author} {\bibfnamefont {P.}~\bibnamefont {Laczkowski}},
  \bibinfo {author} {\bibfnamefont {W.}~\bibnamefont {Savero}}, \bibinfo
  {author} {\bibfnamefont {J.-P.}\ \bibnamefont {Attane}}, \bibinfo {author}
  {\bibfnamefont {C.}~\bibnamefont {Deranlot}}, \bibinfo {author}
  {\bibfnamefont {M.}~\bibnamefont {Jamet}}, \bibinfo {author} {\bibfnamefont
  {J.-M.}\ \bibnamefont {George}}, \bibinfo {author} {\bibfnamefont
  {L.}~\bibnamefont {Vila}}, \ and\ \bibinfo {author} {\bibfnamefont
  {H.}~\bibnamefont {Jaffres}},\ }\href@noop {} {\bibfield  {journal} {\bibinfo
   {journal} {Phys. Rev. Lett.}\ }\textbf {\bibinfo {volume} {112}},\ \bibinfo
  {pages} {106602} (\bibinfo {year} {2014})}\BibitemShut {NoStop}%
\bibitem [{\citenamefont {Gil}\ \emph {et~al.}(2005)\citenamefont {Gil},
  \citenamefont {G{\"o}rlitz}, \citenamefont {Horisberger},\ and\ \citenamefont
  {K{\"o}tzler}}]{gil2005magnetoresistance}%
  \BibitemOpen
  \bibfield  {author} {\bibinfo {author} {\bibfnamefont {W.}~\bibnamefont
  {Gil}}, \bibinfo {author} {\bibfnamefont {D.}~\bibnamefont {G{\"o}rlitz}},
  \bibinfo {author} {\bibfnamefont {M.}~\bibnamefont {Horisberger}}, \ and\
  \bibinfo {author} {\bibfnamefont {J.}~\bibnamefont {K{\"o}tzler}},\
  }\href@noop {} {\bibfield  {journal} {\bibinfo  {journal} {Phys. Rev. B}\
  }\textbf {\bibinfo {volume} {72}},\ \bibinfo {pages} {134401} (\bibinfo
  {year} {2005})}\BibitemShut {NoStop}%
\bibitem [{\citenamefont {Xu}\ \emph {et~al.}(2010)\citenamefont {Xu},
  \citenamefont {Zhang}, \citenamefont {Liu}, \citenamefont {Wang},
  \citenamefont {Li}, \citenamefont {Wu}, \citenamefont {Yu},\ and\
  \citenamefont {Zhang}}]{xu2010anomalous}%
  \BibitemOpen
  \bibfield  {author} {\bibinfo {author} {\bibfnamefont {W.}~\bibnamefont
  {Xu}}, \bibinfo {author} {\bibfnamefont {B.}~\bibnamefont {Zhang}}, \bibinfo
  {author} {\bibfnamefont {Z.}~\bibnamefont {Liu}}, \bibinfo {author}
  {\bibfnamefont {Z.}~\bibnamefont {Wang}}, \bibinfo {author} {\bibfnamefont
  {W.}~\bibnamefont {Li}}, \bibinfo {author} {\bibfnamefont {Z.}~\bibnamefont
  {Wu}}, \bibinfo {author} {\bibfnamefont {R.}~\bibnamefont {Yu}}, \ and\
  \bibinfo {author} {\bibfnamefont {X.}~\bibnamefont {Zhang}},\ }\href@noop {}
  {\bibfield  {journal} {\bibinfo  {journal} {Europhys. Lett.}\ }\textbf
  {\bibinfo {volume} {90}},\ \bibinfo {pages} {27004} (\bibinfo {year}
  {2010})}\BibitemShut {NoStop}%
\bibitem [{\citenamefont {Xu}\ \emph {et~al.}(2008)\citenamefont {Xu},
  \citenamefont {Zhang}, \citenamefont {Wang}, \citenamefont {Chu},
  \citenamefont {Li}, \citenamefont {Wu}, \citenamefont {Yu},\ and\
  \citenamefont {Zhang}}]{xu2008scaling}%
  \BibitemOpen
  \bibfield  {author} {\bibinfo {author} {\bibfnamefont {W.}~\bibnamefont
  {Xu}}, \bibinfo {author} {\bibfnamefont {B.}~\bibnamefont {Zhang}}, \bibinfo
  {author} {\bibfnamefont {Z.}~\bibnamefont {Wang}}, \bibinfo {author}
  {\bibfnamefont {S.}~\bibnamefont {Chu}}, \bibinfo {author} {\bibfnamefont
  {W.}~\bibnamefont {Li}}, \bibinfo {author} {\bibfnamefont {Z.}~\bibnamefont
  {Wu}}, \bibinfo {author} {\bibfnamefont {R.}~\bibnamefont {Yu}}, \ and\
  \bibinfo {author} {\bibfnamefont {X.}~\bibnamefont {Zhang}},\ }\href@noop {}
  {\bibfield  {journal} {\bibinfo  {journal} {Eur. Phys. J. B}\ }\textbf
  {\bibinfo {volume} {65}},\ \bibinfo {pages} {233} (\bibinfo {year}
  {2008})}\BibitemShut {NoStop}%
\bibitem [{\citenamefont {V\'elez}\ \emph {et~al.}(2016)\citenamefont
  {V\'elez}, \citenamefont {Golovach}, \citenamefont {Bedoya-Pinto},
  \citenamefont {Isasa}, \citenamefont {Sagasta}, \citenamefont {Abadia},
  \citenamefont {Rogero}, \citenamefont {Hueso}, \citenamefont {Bergeret},\
  and\ \citenamefont {Casanova}}]{PhysRevLett.116.016603}%
  \BibitemOpen
  \bibfield  {author} {\bibinfo {author} {\bibfnamefont {S.}~\bibnamefont
  {V\'elez}}, \bibinfo {author} {\bibfnamefont {V.~N.}\ \bibnamefont
  {Golovach}}, \bibinfo {author} {\bibfnamefont {A.}~\bibnamefont
  {Bedoya-Pinto}}, \bibinfo {author} {\bibfnamefont {M.}~\bibnamefont {Isasa}},
  \bibinfo {author} {\bibfnamefont {E.}~\bibnamefont {Sagasta}}, \bibinfo
  {author} {\bibfnamefont {M.}~\bibnamefont {Abadia}}, \bibinfo {author}
  {\bibfnamefont {C.}~\bibnamefont {Rogero}}, \bibinfo {author} {\bibfnamefont
  {L.~E.}\ \bibnamefont {Hueso}}, \bibinfo {author} {\bibfnamefont {F.~S.}\
  \bibnamefont {Bergeret}}, \ and\ \bibinfo {author} {\bibfnamefont
  {F.}~\bibnamefont {Casanova}},\ }\href@noop {} {\bibfield  {journal}
  {\bibinfo  {journal} {Phys. Rev. Lett.}\ }\textbf {\bibinfo {volume} {116}},\
  \bibinfo {pages} {016603} (\bibinfo {year} {2016})}\BibitemShut {NoStop}%
\end{thebibliography}

%

\clearpage

\begin{figure}[tb]
\includegraphics[scale=1]{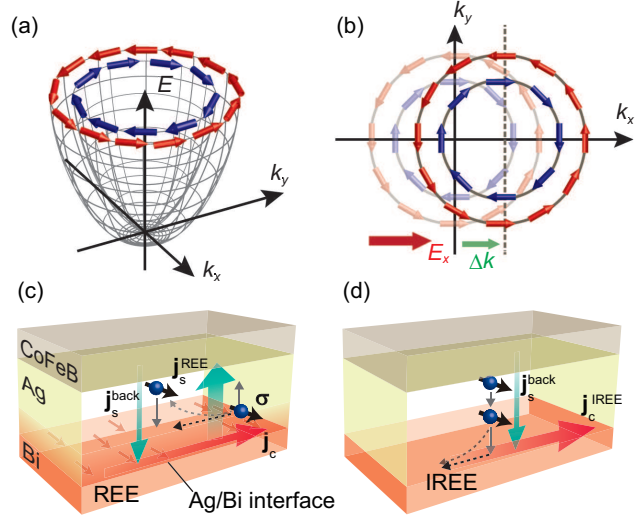}
\caption{(a) The spin-split dispersion of a Rashba 2DEG system. (b) Fermi Contours under an external electric field $E_x$. A shift $\Delta k$ of the Fermi circles gives rise to a spin accumulation. (c) A schematic illustration of the REE in the Bi/Ag/CoFeB trilayer. The REE generates a spin current ${\bf j}_\text{s}^\text{REE}$ from a charge current ${\bf j}_\text{c}$. The spin current is reflected at the Ag/CoFeB interface, generating a back flow spin current ${\bf j}_\text{s}^\text{back}$. (d) A schematic illustration of the IREE, which converts ${\bf j}_\text{s}^\text{back}$ into an additional charge current ${\bf j}_\text{c}^\text{IREE}$. 
}
\label{fig1} 
\end{figure}

\clearpage

\begin{figure}[tb]
\includegraphics[scale=1]{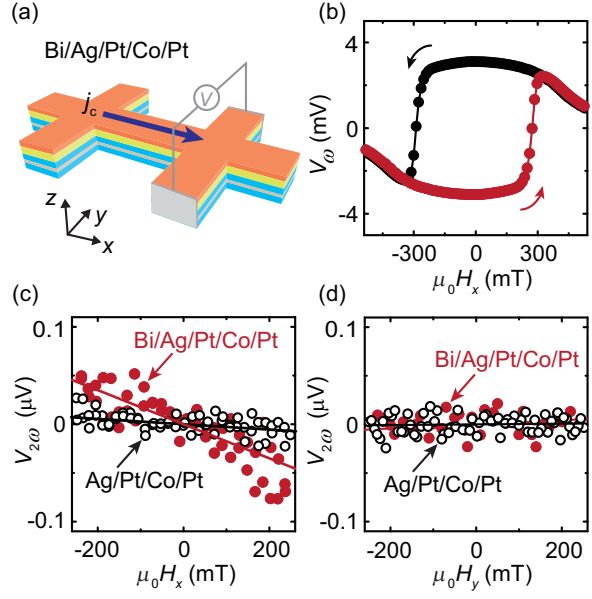}
\caption{(a) A schematic illustration of the Bi/Ag/Pt/Co/Pt multilayer. The arrow in blue represents the applied charge current $j_\text{c}$. (b) The first harmonic voltage $V_\omega$ as a function of the magnetic field $\mu_0 H_x$ applied along the $x$ axis for the Bi/Ag/Pt/Co/Pt multilayer. (c) $\mu_0 H_x$ dependence of the second harmonic voltage $V_{2 \omega}$ for the Bi/Ag/Pt/Co/Pt (the solid red circles) and Ag/Pt/Co/Pt (the open circles) multilayers. The solid lines are the linear fit to the data. (d) The second harmonic voltage $V_{2\omega}$ as a function of the magnetic field $\mu_0  H_y$ applied along the $y$ axis for the Bi/Ag/Pt/Co/Pt (the solid red circles) and Ag/Pt/Co/Pt (the open circles) multilayers.}
\label{fig2} 
\end{figure}

\clearpage

\begin{figure}[tb]
\includegraphics[scale=1]{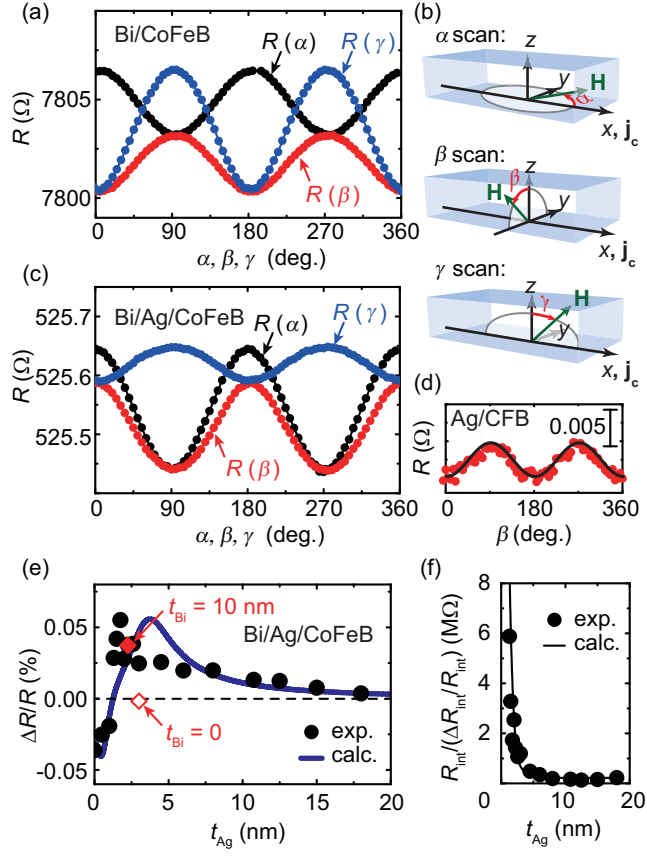}
\caption{The longitudinal resistance $R$ for (a) the Bi/CoFeB bilayer and (c) the Bi/Ag/CoFeB trilayer as a function of the rotation of the magnetic field $\mu_0 H=10$ T. The magnetic field angle $\alpha$, $\beta$, and $\gamma$ are defined in (b). (d) The $\beta$ dependence of $R$ for the Ag/CoFeB bilayer. The solid curve is a function proportional to $\sin^2\beta$. (e) The Ag-layer thickness $t_\text{Ag}$ dependence of $\Delta R/R$ for the Bi/Ag/CoFeB trilayer. The solid circles are the experimental data and the solid curve is the result of the calculation. The values of $\Delta R/R$ for the Bi($t_\text{Bi}$)/Ag/CoFeB trilayer are also shown, where $t_\text{Bi}$ is the thickness of the Bi layer. (f) The solid circles are $t_\text{Ag}$ dependence of $R_\text{int}/(\Delta R_\text{int}/R_\text{int})$ extracted from the $t_\text{Ag}$ dependence of $\Delta R/R$ using Eq.~(\ref{thicknessdep}). The solid curve is $R_\text{int}/(\Delta R_\text{int}/R_\text{int})=A\exp(-t_\text{Ag}/B)+C$, where $A$ = 88 M$\Omega$, $B$ = 0.55 nm, and $C$ = 0.22 M$\Omega$.
}
\label{fig3} 
\end{figure}

\clearpage

\begin{figure}[tb]
\includegraphics[scale=1]{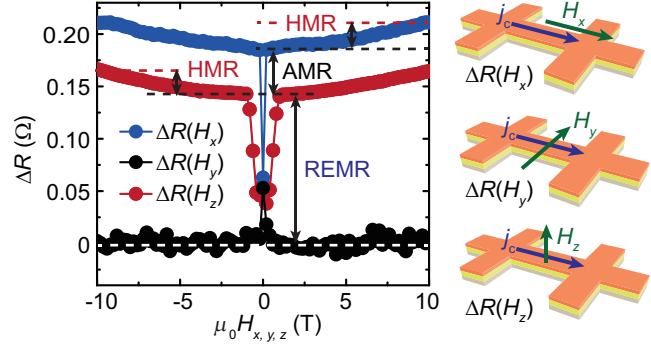}
\caption{The change of the longitudinal resistance $\Delta R(H_{x,y,z})=R(H_{x,y,z})-R_0$ plotted as a function of the magnetic field oriented along the $x$ axis (blue), $y$ axis (black), and $z$ axis (red) for the Bi/Ag/CoFeB trilayer. $R_0$ is the resistance measured at $\mu_0 H_y=10$ T.}
\label{fig4} 
\end{figure}

\end{document}